# Predicting the Performance of a Future United Kingdom Grid and Wind Fleet When Providing Power to a Fleet of Battery Electric Vehicles

Anthony D Stephens[a] and David R Walwyn[b]
Correspondence to tonystephensgigg@gmail.com

**Abstract**

Sales of new petrol and diesel passenger vehicles may not be permitted in the United Kingdom (UK) post-2030. Should this happen, it is likely that vehicles presently powered by hydrocarbons will be progressively replaced by Battery Electric Vehicles (BEVs). This paper describes the use of mathematical modelling, drawing on real time records of the UK electricity grid, to investigate the likely performance of the grid when supplying power to a fleet of up to 35 million BEVs.

The model highlights the importance of understanding how the grid will cope when powering a BEV fleet under conditions similar to those experienced during an extended wind lull during the 3$^{rd}$ week of January 2017. Allowing a two-way flow of electricity between the BEVs and the grid, known as the vehicle-to-grid (V2G) configuration, turns out to be of key importance in minimising the need for additional gas turbine generation or energy storage during wind lulls. This study has shown that with the use of V2G, it should be possible to provide power to about 15 million BEVs with the gas turbine capacity currently available. Without V2G, it is likely that the current capacity of the gas turbines and associated gas infrastructure might be overwhelmed by even a relatively small BEV fleet.

Since it is anticipated that 80% of BEV owners will be able to park the vehicles at their residences, widespread V2G will enable both the powering of residences when supply from the grid is constrained and the charging of BEVs when supply is in excess. The model shows that this configuration will maintain a constant load on the grid and avoid the use of either expensive alternative storage or hydrogen obtained by reforming methane. There should be no insuperable problem in providing power to the 20% of BEV owners who do not have parking at their residences; their power could come directly from the grid.

---

[a] Corresponding author
[b] Department of Engineering and Technology Management, University of Pretoria, South Africa

## Introduction

Some regard wind power as a transformational technology which will enable Battery Electric Vehicles (BEVs) to replace petrol and diesel cars, provide hydrogen to heat homes, make steel without the need for hydrocarbons, and generally enable the United Kingdom (UK) to decarbonise its way of life. The Prime Minister, a one-time wind adversary, is clearly a recent convert to this view, telling his party conference on 6th October 2020, when announcing an increase in the capacity of the UK wind fleet to 40 GWc, that:

> "Yes, you heard me right. Your kettle, your washing machine, your cooker, your heating, your plug-in electric vehicle – the whole lot of them will get their juice cleanly and without guilt from the breezes that blow around these islands". (Elliott, 2020)

The problem with this attractive sounding suggestion is that it is seriously deficient in two respects. Firstly, the additional wind generation announced would only meet a very small fraction of the Prime Minister's promised benefits. Secondly, he chose to ignore the inconvenient truth that, even with a highly distributed wind fleet, there are times when the UK experiences practically no wind generation, sometimes for lengthy periods. This would have serious consequences for any system which had become over- reliant on wind generation.

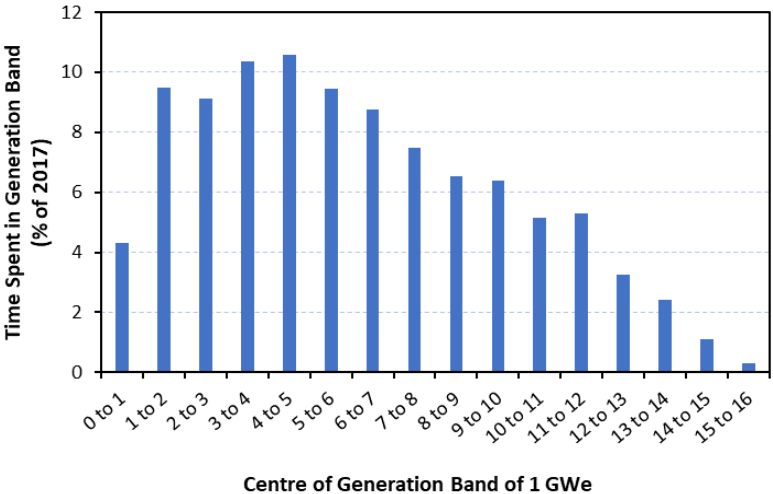

**Figure 1. Percentage of time spent by the wind fleet in generation bands of 1 GWe (2017)**

Figure 1 is a wind histogram which shows the percentage of time the wind fleet spent in generation bands of 1 GWe in 2017, when the nominal wind fleet capacity was 20 GWc [c]. It may be seen that for over 4% of the year the wind fleet generated between 0 and 1 GWe. One of the important issues in this study has been gaining understanding of the likely performance of a future grid powering a BEV fleet during wind lulls. From the electricity grid's point of view the most serious wind lulls occur during the winter months when demands on the grid are at their height. As the National Grid Director Fintan Slye told the Times on 9th November 2020, when giving a rare warning about possible imminent electricity shortages, that:

> "the thing you need to worry about is low wind scenarios on a cold winter's day, mid-December or mid-January, really still high pressure sitting atop the UK". (Gosden, 2020a)

The incentive to decarbonise UK transport may be seen in Table 1 which summarises the progress in reducing UK carbon emissions between 1990 and 2017. The biggest reduction in carbon emission, from

---

[c] Throughout this article GWe and MWe are used to refer to the gigawatts and megawatts respectively of generated power. GWc on the other hand refers to gigawatts of generation capacity.

242.1 to 105 metric tonne (MT) per annum (p.a.), was in energy supply, although most of this reduction was the result of moving from coal fired to gas fired generation and reducing demand on the grid as a consequence of manufacturing industry being relocated to countries such as China (Li and Hewitt, 2008). Of the reduction in energy supply emissions of 137.1 MT p.a., only 29 MT p.a. can be attributed to UK's investment in its 20 GWc wind fleet. The wind fleet generated around 6 GWe in 2017, thus displacing 6 GW of gas generation with a carbon emissions content of 4.8 MT p.a. per GWe of gas generation.

**Table 1. UK carbon emissions MT p.a. for 1990 and 2017**

|  | 1990 | 2017 |
|---|---|---|
| Energy Supply | 242.1 | 105.0 |
| Business | 111.9 | 65.8 |
| Transport | 125.3 | 124.4 |
| Residential | 78.4 | 64.1 |
| Other | 36.4 | 7.4 |
| **Total** | **594.1** | **366.7** |

Source: UK government records

Transport was the sector most resistant to improvement, contributing 124.4 MT of emissions in 2017 to the total emissions of 366.9 MT p.a., with little change since 1990. Approximately 35 million petrol and diesel passenger vehicles then generated 66.3 MT p.a. of carbon, making a switch to BEVs one of the most promising ways of reducing carbon emissions. The paper will explore the possibility of using wind to power a fleet of up to 35 million vehicles. Of course, BEVs come in all shapes and sizes. For modelling purposes, we shall assume a mid-range BEV, travelling 50 km per day with an average daily energy consumption of 10 kWh and a total energy storage capacity of 30 kWh. The average power consumption of the fleet would then be 14.6 GW and the stored energy, if all the vehicles were to be fully charged, would be 1,050 GWh.

**Modelling the Wind Fleet**

The demand on the grid, and the total wind/solar generation during week 17 of 2017 are shown in Figure 2. The sources of generation were recorded every 5 minutes so the records for a week comprise 2016 data sets. Only around ⅔$^{rd}$ of wind generation is recorded by the grid, the other ⅓$^{rd}$ being "embedded" and unseen. However, government records allow the capacity of the "embedded" component to be estimated, so that the recorded generation may be scaled to make an allowance for the "embedded" component. The wind fleet capacity factor, the average annual generation divided by the nominal wind fleet capacity, varies from year to year, having in recent years been generally in the range 30 ± 3 %. To put the records for different years on the same basis, the wind generation records are scaled to ensure a capacity factor of 30%.

A number of generation sources which normally do not vary greatly during a year may be represented by a single composite variable which we shall call Base Generation. The main components of Base Generation, which totalled around 13 GWe in 2017 were nuclear generation (c 8 GWe), imports (c 3 GWe) and bio generation (c 2 GWe). However, will be circumstances when Base Generation might be significantly different from this figure, so it must be treated as a variable in the model.

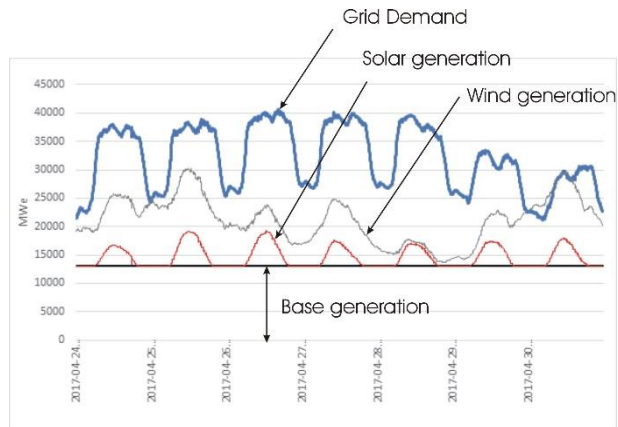

**Figure 2. Grid records for week 17 (2017) showing the contributions of Base, wind and solar generation to Grid Demand**

Source: Gridwatch (2017)

In the representation of Figure 2, solar generation is shown stacked above Base Generation, and wind generation stacked about solar generation. The area between Base Generation and Grid Demand was traditionally the operational area of coal and gas generation, but wind and solar generation are now given preferential access. The National Grid announced in April 2020 that a combination of low Grid Demand and increasing wind/solar generation capacities was beginning to lead to occasions when wind farms were having to be paid to curtail their output (Jillian Ambrose, 2020). An occasion when wind generation was more than sufficient to satisfy Grid Demand may be seen in Figure 2 during the early hours of 30$^{th}$ April. Two of the reasons for choosing 2017 records for this study, rather than the records of earlier or later years, were that wind curtailment would have increased in later years, and 2017 was the first year for which solar generation records are available on the grid (Stephens and Walwyn, 2020). Solar generation must be included in any model since wind and solar generation share the same operational area.

In 2014, the Royal Academy of Engineering reported that consultants working for the Department of Energy and Climate Change had estimated the likely future upper limit of the UK wind fleet to be between 34 GWc and 75 GWc (Royal Academy of Engineering, 2014). In this study we shall consider a range of wind fleets up to 80 GWc. Figure 3 shows the predictions, by extrapolation from the 2017 records, of wind generation during week 17 of 2017 had the wind fleet size been 20 GWc (as in 2017), 40 GWc, 60 GWc or 80 GWc.

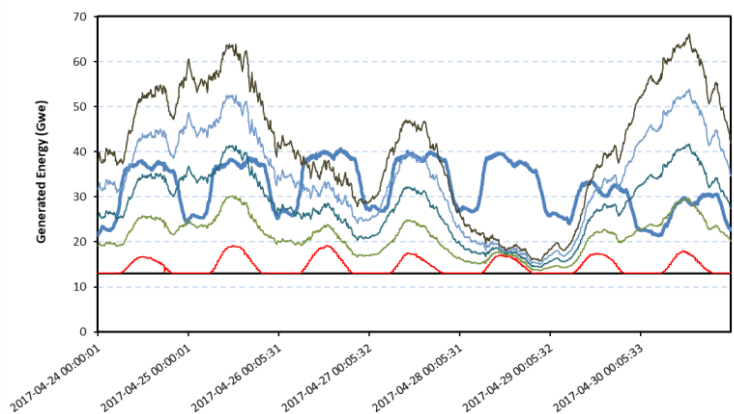

**Figure 3. Grid representation for week 17 of 2017 the predicted output of wind fleets ranging in size from 20 GWc to 80 GWc**

On April 28th 2017 there would have been an energy deficit of 338 GWh, even with a wind fleet of 80 GWc, while on 30th April there would have been a surplus of 720 GWh. It is sometimes suggested that the problem of such enormous grid surpluses/ deficits may be solved by storing surpluses for later use. However, although energy storage systems are being increasingly deployed, they are on a very small scale relative to total demand (Stephens and Walwyn, 2017). By the end of 2020, the world's largest energy storage system, excluding pumped storage and molten salt systems, was reportedly the 800 MWh vanadium redox flow battery in China, built by Rongke Power and its partner, US UniEnergy Technologies, although the status of this project is unclear (Norge Mining, 2020). The largest operating facility is the 230 MWh facility lithium ion battery in California, operated by Gateway Energy Storage (LS Power, 2020). An array sufficiently large to mitigate the wind lull of 28th April 2017 would have needed 300 such facilities, at a capital cost of the order of £60 Bn. It must also be remembered that 5-day wind lulls are not infrequent (there was a 10-day wind lull in January 2017 which will be discussed later) and wind surpluses/deficits do not present themselves in an orderly fashion.

The extremely high-cost facilities needed to mitigate wind lulls would have a very low utilisation. The current small-scale energy storage systems do not aim to resolve grid sized problems, but to capitalise on the highly variable price of electricity -buying when the price is low and selling back to the grid when the price is high. Gresham House Storage is the largest UK storage company trading in this way, but has to date only invested some £200M in storage ventures ranging from 5 MW to 40 MW (stored energy unknown) (Hosking, 2020). The only electrical storage facility to have similar stored energy as the deficits/ surpluses discussed above is likely to be that of a future fleet of BEVs. As we shall discuss later, the BEV storage energy will not be used to mitigate wind deficits/ surpluses, but to match the varying electrical demands of grid and BEV fleet. The conclusion, which has also been reached by others, is that wind generation which exceeds what the grid is able to accommodate will continue to be curtailed (Siddique and Thakur, 2020; Barnhart, Dale, Brandt and Benson, 2013; Marcacci, 2013). The model which generated Figure 3 was modified to reflect such curtailment, the results being shown in Figure 4.

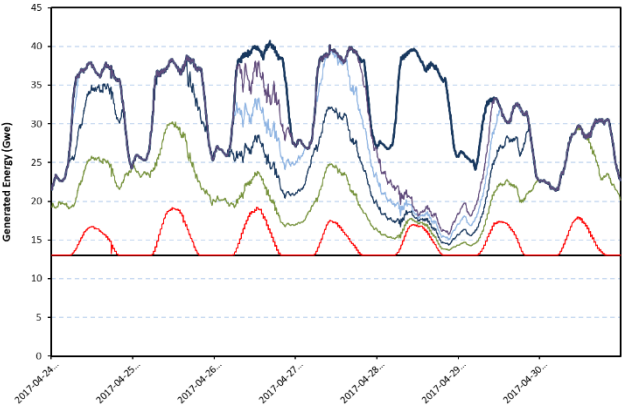

**Figure 4. Predictions for week 17 of 2017 for wind fleets up to 80 GWc in size with wind generation which exceeds Grid Demand being curtailed**

Although the model which generated Figures 3 and 4 did so for all wind fleet sizes from 20 GWc to 80GWC in steps of 10 GWc, only the wind generation predictions for wind fleet in steps of 20 GWc are shown in the interest of clarity. The model calculates the weekly averages for each sized wind fleet and the averages for the 52 weeks of the year are combined to produce annual average wind generation predictions, as shown in Figure 5. What this shows is that up to 20 GWc, the wind fleet generates 0.3 GWe (average efficiency of 30%) for every additional GWc of wind fleet capacity, but above 20 GWc there is a progressive decrease in efficiency because of increased wind curtailment. This is consistent with the National Grid first announcing the need for wind curtailment in 2017 when the wind capacity was 20 GWc (Gosden, 2017).

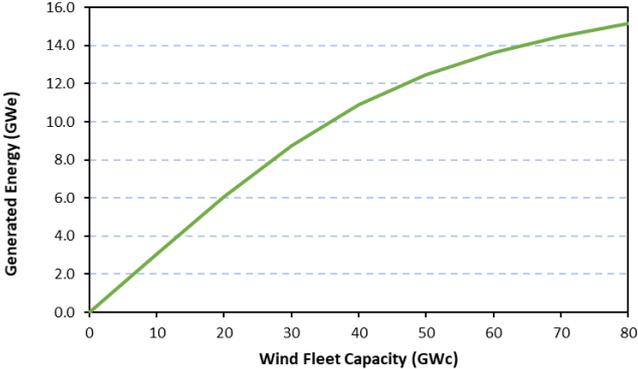

**Figure 5. Wind fleet characteristic for grid configuration as in 2017**

Future grid configurations are likely to have very different configurations to those in 2017, and a model of wider applicability will need to be able to take account of different cyclic components of Grid Demand, different levels of Base Generation and different levels of average Grid Demand.

The authors have previously reported that removing the cyclic component entirely, by replacing the real time Grid Demand with average Grid Demand, had only a minimal effect on the calculated wind fleet efficiency (Stephens and Walwyn, 2018a). An explanation for this initially surprising finding is that for half the year the removal of the cyclic component reduces the calculation of wind curtailment and for the other half increases it, the two effects cancelling each other. As far as the wind fleet is concerned, its performance is determined by any combination of average Grid Demand and Base Generation which leads to its operational area being as shown in Figure 6. The size of the operational area is determined by the dimension Hdrm, and this recognition leads to a very simple model representation which has general applicability.

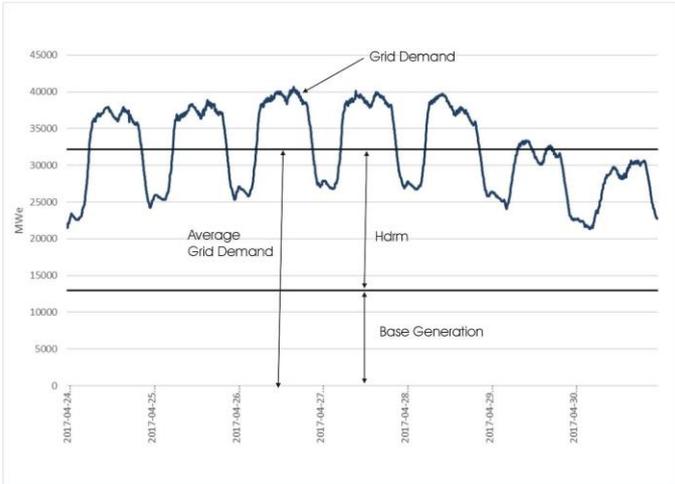

**Figure 6. The concept of the wind fleet's Hdrm**

Since a 1 GWe change in Grid Demand has the same effect on the operational area as 1 GWe change in Base Generation, a measure of the size of the operational area is the difference between the two, which we shall call the wind fleets Headroom or Hdrm, where:

$$Hdrm = Average\ Grid\ Demand - Base\ Generation$$

In 2017 average Grid Demand was 33 GWe and Base Generation was 13 GWe, so Hdrm was 20 GWe, and Figure 5 is valid for all combinations of Base Generation and Grid Demand with a Hdrm of 20 GWe. The important corollary is that by choosing Base Generation values judiciously we are able to generate families of GWe vs GWc curves for particular Hdrm values. Thus, if the 2017 weekly data are run with Base Generation = 3 GWe, rather than 13 GWe, the resulting GWe vs GWc curve will be for a Hdrm of 30 GW. Figure 7 shows the families of GWe vs GWc predictions for Hdrm values of 20, 25, 30 and 35 using 2017 records (red dotted curves), together with the predictions generated previously in the same way using 2014 grid records.

Bearing in mind that it was necessary to make some engineering approximations in order to describe the grid configurations in terms of Hdrm only, it gives some confidence in the use of only a single year's records to produce a model of general application, that there is so little spread in the characteristic curve predictions using the records of 2014 to 2017; a period when the wind fleet increased in size by around 50%. As the authors explain elsewhere (Stephens and Walwyn, 2018b), the reason for the records of different years producing such similar characteristic curves is that the wind histograms for different years are similar. Although the wind at any time is random, the annual histograms reveal similar variability in different years. If this were not the case, the characteristic curves produced using different year's records would be different. In fact, it is possible to produce close approximations to the characteristic curves of Figure 7 directly from the annual wind generation histograms (Stephens and Walwyn, 2018b). Unfortunately, because the wind and solar fleets share the same operational area, the two methods of producing the characteristic curves will diverge as the solar fleet increases in size. The more computationally complex method of averaging the real time records will produce the more accurate curves.

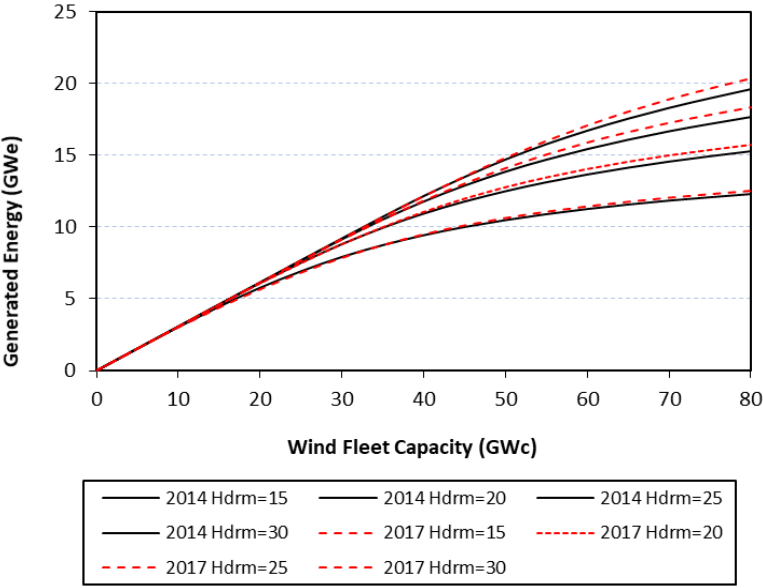

**Figure 7. Wind fleet performance predictions for different values of Hdrm derived using 2014 and 2017 records**

Two factors not considered in the characteristic curves of Figure 7 are future changes in solar generating capacity and in the wind fleet's capacity factor (i.e. average generation divided by wind fleet capacity) which will result from a higher percentage of generation being off-shore. However, the model includes scaling factors to allow these factors to be considered. It will still be possible to use the 2017 records to generate characteristic equations appropriate to the new factors, each week's records being adjusted using the new scaling factors, and the weekly predictions averaged as described previously. (The solar generation shown in Figures 2, 3 and 8 are those recorded in 2017 but, because the solar generation capacity has increased considerably since 2017, twice the solar generation of 2017 was used in calculating the annualised Figure 5, 7 and 12).

**Modelling the Grid When Powering a Battery Electric Vehicle Fleet**

As evidenced by the National Grid's warning of possible power shortages on 9[th] November 2020 (Gosden, 2020a), even though Grid Demand was low and the temperature was 10°C, the National Grid was already struggling to meet peak demands. Should all the UK's 35 million petrol and diesel passenger vehicles be replaced by BEVs, as is the UK government's current intention, this would increase average demand by around 50%. This would be extremely demanding, particularly during wind lulls. However, a large fleet of BEVs would have a significant amount of stored energy which, if properly managed, could be used to reduce peak demand on the grid. In this section we shall discuss the optimal BEV charging pattern, and how the model described earlier may be modified to investigate how the BEVs' stored energy may be used to minimise the impact of the BEVs on the grid.

The optimal arrangement for the grid would then be for the charging system to incentivise vehicle owners to take their charge in such a way that a level demand during the week is achieved. This level demand would, by definition, be the average Grid Demand that week plus average BEV demand. A model was described earlier in the paper in which wind generation was curtailed at the level of Grid Demand, resulting in the predictions as in Figure 4. This model was modified to curtail wind generation at average Grid Demand plus average BEV demand. Figure 8 shows the model prediction for week 17 of 2017 had there been a BEV fleet of 35 million vehicles charged in the optimal manner described above. Average Grid Demand that week was 32.1 GWe and, with the addition BEV demand of 14.6 GWe, total demand would have been 46.7 GWe.

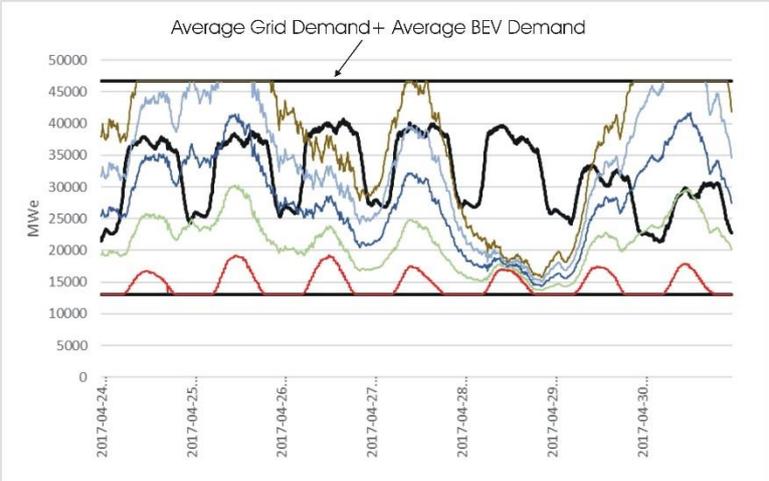

**Figure 8. Wind generation limited at average Grid Demand plus average BEV demand for week 17 of 2017 (with a BEV fleet of 35 million vehicles)**

The model allows a calculation to be made of the phasing of the BEV demand on the grid which would be needed to level the total demand, as shown in Figure 8. By definition, this must be the difference between the total demand and the real time Grid Demand, as shown in Figure 9.

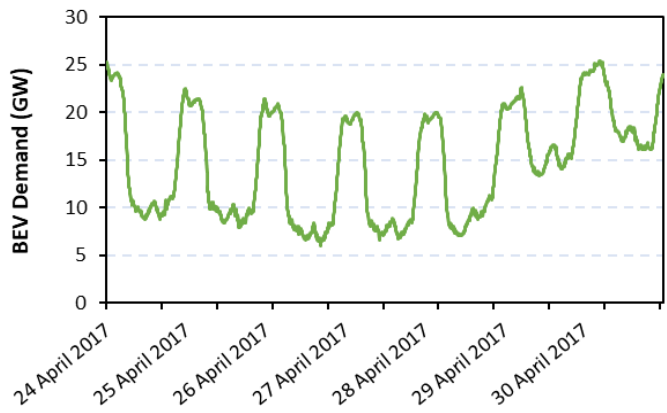

**Figure 9. Calculated BEV demand from a 35 million BEV fleet to create a level total demand on the grid during week 17 of 2017**

We shall assume a very simple model of the consumption of the BEV fleet at night, 9pm to 6am, being 20% of that during the day, 6am to 9pm, as shown in Figure 10.

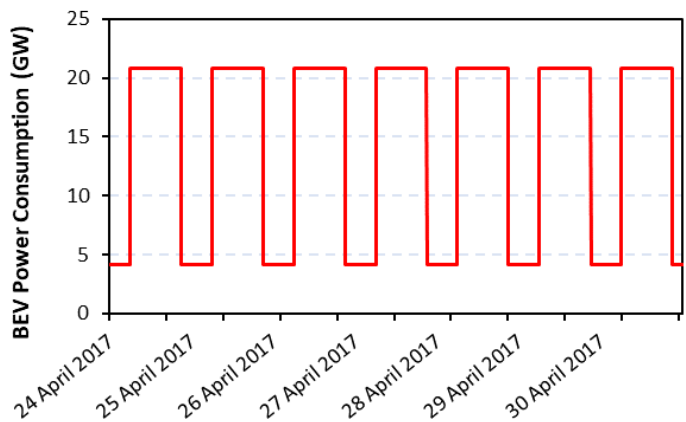

**Figure 10. Assumed power consumption of the BEV fleet during the week**

The fleet of 35 million BEVs with an average stored energy of 30 kWh parameters would have the potential to store 1,050 GWh. Assuming that 80% of that energy was stored in the BEV fleet at the beginning of the week, the stored energy trajectory during the week would have been as shown in Figure 11. The simulation reveals that the stored energy in a 35 million BEV fleet should be more than sufficient to both power the BEVs to the satisfaction of the vehicle owners and phase the demand on the grid in such a way that total demand during the week is constant.

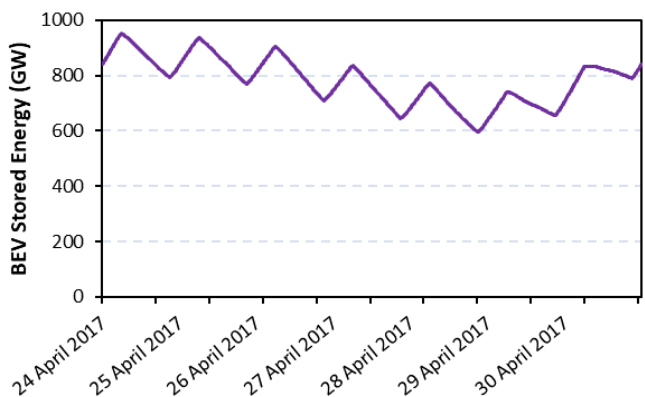

**Figure 11. BEV stored energy trajectory during week 17 of 2017**

Earlier in the paper it was explained how the data for 52 weeks, for which wind generation was limited by Grid Demand, were averaged to produce characteristic curves for different Hdrm values. The 52-week versions of the model, modified to limit wind generation at average Grid Demand plus average BEV Demand, are similarly averaged to produce the families of GWe vs GWc characteristic curves for different sized BEV fleets. These are shown in Figure 12, the lower orange curve being for no BEV fleet and the curves above being for fleets ranging from 15 million to 35 million BEVs in size.

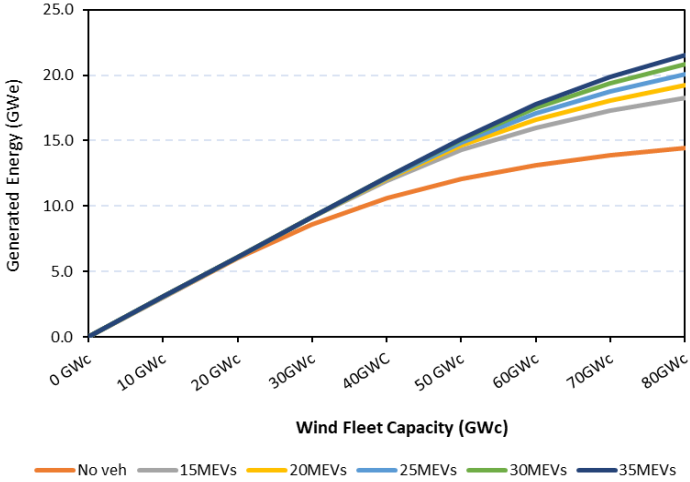

**Figure 12. Family of GWe vs GWc characteristic curves for different sized BEV fleets**

The data underpinning Figure 12 enables the sizes of wind fleets needed to power various sized BEV fleets to be calculated. The results are summarised in Table 2, together with calculations of the electrical storage capacity of the BEV fleets, the MT p.a. of carbon emissions saved and the cost of the BEV batteries assuming a unit cost of Euro 255 per kWh.

**Table 2. Calculated wind fleet sizes to power various sizes of BEV fleets[d]**

| Size of BEV fleet in million | 15 | 20 | 25 | 30 | 35 |
|---|---|---|---|---|---|
| Average power required by BEV fleet (GWe) | 6.2 | 8.3 | 10.4 | 12.5 | 14.6 |
| Size of wind fleet required (GWc) | 41.8 | 49.3 | 57.5 | 66 | 75 |
| Electrical storage capacity of BEV fleet (GWh) | 450 | 600 | 750 | 900 | 1050 |
| Reduction in carbon emissions (MT p.a.) | 28.4 | 37.9 | 47.4 | 56.8 | 66.3 |
| Cost of BEV batteries (Euro Bn) | 115 | 153 | 191 | 229 | 268 |

Table 2 includes a typical manufacturers' quoted cost for a Li ion battery of Euro 155 per kWh (Henze, 2020), to which has been added Euro 100 per kWh for the charger which we shall later argue will be essential if the operation of the grid is to be optimised. Table 2 suggests that it should be possible to power a 35 million BEV fleet by increasing the wind fleet to 75 GWc and, in so doing, save 66.3 MT p.a. of carbon emissions. However, these predictions are based on steady state average assumptions. In the next section we shall consider whether it is likely that it will be possible to achieve these objectives in practice.

---

[d] It is noted that the power required is in addition to the 6 GWe generated by the 20 GWc wind fleet

**The Problem of Wind Lulls**

Wind lulls pose significant problems for systems which rely on wind power, and this will be particularly the case for a grid powering a BEV fleet in addition to its normal demand. It will be necessary to have a shadow capacity of dispatchable source of generation to provide power when called upon to do so. In practice, this will be a shadow fleet of combined cycle gas turbines (CCGTs). An important question is whether it is likely that there will be sufficient standby gas turbine capacity for a grid powering a BEV fleet while meeting the Secretary of States Level of Load Expectation standard of the grid having no more than 3 hrs of outage a year (National Grid, 2020).

The model allows a simulation to be made of the likely performance of the grid with a BEV load during week 3 of 2017, January 16th to 23rd, one of the most severe tests of the grid in recent years. This period was part of a 10-day event during which there was a stationary high-pressure area over much of Europe (see Figure 13 for wind synoptic on 23rd January) and the UK, German and French grids recorded minimal wind generation. Despite France having 63 GWe of nuclear capacity, its domestic heating is mainly electrical, and it had no surplus to export to the UK and other countries due to the mid-winter conditions in the country. The interconnectors with France and Holland normally provide the UK with around 3 GWe of imports, but the records for the week show no net imports to the UK.

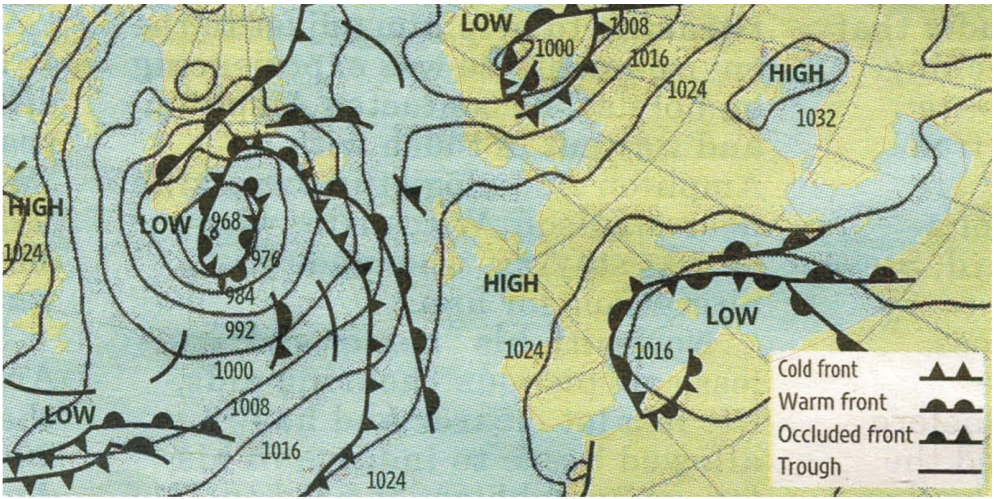

Figure 13. Weather synoptic for 23rd January 2017

Most of the UK's existing nuclear capacity is likely to be retired by the end of the 2020s(HM Government, 2020), and a best estimate of nuclear capacity in 2030 is perhaps 4 GWe. Although Base Generation was around 13 GWe for most of 2017, for the simulation of future scenarios during winter wind lulls, it is assumed to be only 7 GWe, to take account of both reduced nuclear capacity and unavailability of imports to the UK.

During week 3 of 2017 peak Grid Demand was 50.9 GWe average Grid Demand was 41.1 GWe and Figure 14 shows the simulation results for a grid with a BEV fleet of 35 million vehicles and wind fleets from 20 GWc to 80 GWc. The distance between the constant total demand of 55.6 GWe and the 80 GWc wind generation curve in Figure 14 represents the shortfall which the gas turbines are required to mitigate, and is shown in Figure 15. The peak gas turbine requirement during the week was 47.0 GWe and the average gas turbine generation requirement was 40.1 GWe, making the energy deficit during the week caused by the wind lull a total of 11,551 GWh.

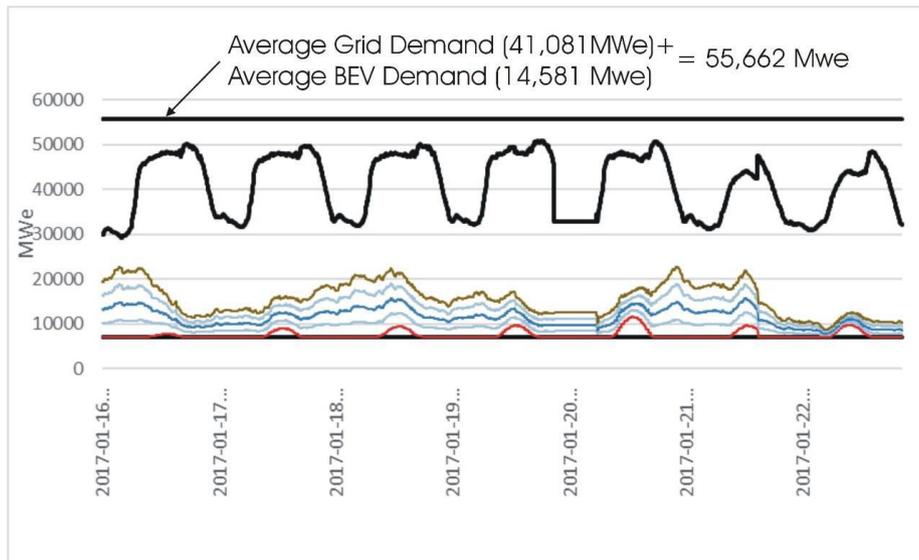

**Figure 14. Model simulation for 3rd week of January 2017 with a 35 million BEV fleet and a Base Generation of 7 GWe, showing very low wind generation predictions from wind fleets of 20 GWc, 40 GWc, 60 GWc and 80 GWc**

It is useful to compare this deficit with the 0.64 GWh of what is likely to be Europe's largest store of energy of 0.64 GWh in 2024 (Gosden, 2020b). This strongly reinforces the conclusion that stored energy will never be able to mitigate wind lulls, and that a wind fleet will always need a shadow fleet of gas turbine generators for this purpose.

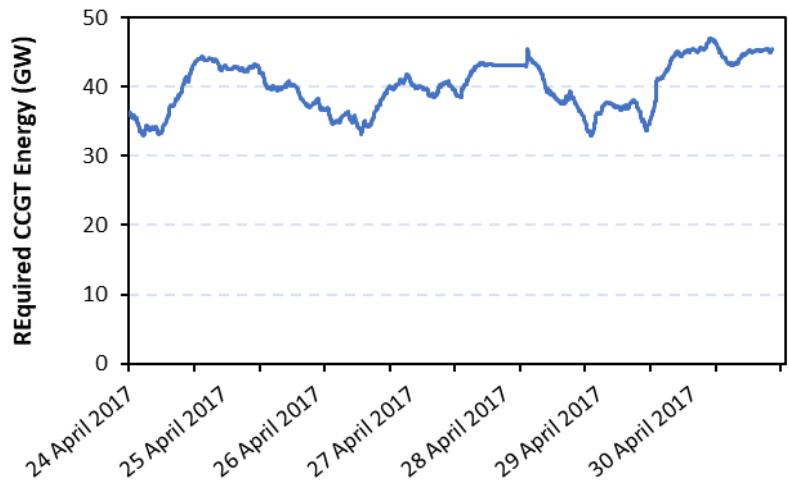

**Figure 15. Gas turbine generation required during week 3 of 2017 for a grid providing power to a 35 million BEV fleet**

**Discussion**

In commenting on the Prime Minister's plan to make Britain greener (Elliott, 2020), the Economist said that

> *"Mr Johnson's plan to get Briton's into electric vehicles is bold. … but boldness is not enough. To avoid a backlash and an embarrassing retreat, it helps to have a plan… to pledge to quadruple power production for offshore power leaves unanswered questions of how the energy market will cope with a surge in intermittent supply".* (The Economist, 2020a)

The National Audit Office mirrored the Economist's criticism of the government for

> *"not having collated the costs and benefits of meeting its net zero target which was likely to cost hundreds of billions of pounds …. The government needs to develop and monitor clear relevant and consistent data on progress on net zero policies across government and gather information on how much it has committed and spent."* (Webster, 2020)

The unease of the Economist and National Audit Office about the Prime Minister's unwillingness to cost any of his projects must be exacerbated by the cost estimates of others. We have seen that a 75 GWc wind fleet would be needed to power a 35 million BEV fleet and the European Commission estimates that it would cost Euro 150 BN to invest in 50 to 75 GW of solar and wind capacity by 2030 (National Grid, 2020).

Surprisingly, there is little mention of the challenge that will be presented to the grid by the addition of several million BEVs in the UK's present energy planning documents. The Energy White Paper, released in December 2020, mentions only in passing the Ovo Energy V2G demonstration study and places the huge undertaking of the required expansion to the charging grid firmly in the hands of the private sector (HM Government, 2020).

The White Paper appears to draw substantially on the National Grid Strategy for 2050, which was published a few months earlier in July 2020 (National Grid, 2020) and reproduces its weaknesses. Neither documents recognise that it is insufficient to consider only the steady state conditions when determining whether or not a strategy, which involves highly variable sources of renewable energy, is a viable one. The 2050 flow chart advises that there will be an annual average generation of electricity of 68.9 GWe, of which 40.3 GWe will be wind generation, but no mention is made of how the grid will cope when the wind fleet fails to produce 40.3 GWe (National Grid, 2020).

The worst-case scenario would be the BEV fleet taking its charge mainly during the day. The peak Grid Demand of 50.9 GWe might then be increased by around 30 GWe, with a total demand of around 81 GWe. With contributions of 7 GWe from Base Generation and only 2 GWe from a 75 GWc wind fleet, we calculate that this would lead to a peak demand on the gas turbines of around 72 GWe, and an average annual utilisation of the gas turbine fleet of only 16%.

The BBC has reported a promising trial carried out by OVO Energy in conjunction with 319 BEV owners who had been provided with chargers at their residences (Lewington, 2020). A considerably lower peak demand on the gas turbines would be achieved if the BEVs were charged using devices which allowed two-way flow of energy between BEV and the grid, known as vehicle to grid (V2G) charging. Figure 14 suggests that, with V2G charging, Grid Demand would have been 55.7 GWe during the wind lull of 3$^{rd}$ January, when charging a 35 million BEV fleet, and Figure 15 suggests a peak demand on the gas turbines of 47.0 GWe. To achieve the constant Grid Demand of Figure 14, it would not be necessary for all 35 million BEVs to be V2G enabled, but only for a sufficient number to provide the required back-up power. The National Grid estimates that by 2050 between 75% and 80% of BEV owners will park their vehicles at their residences and have V2G chargers (National Grid, 2020). The 20% of BEV owners unable to park at their places of residence are likely to charge their vehicles mainly during the day, drawing energy from the grid which, in turn, would be able to draw energy from V2G BEVs when needed. The model suggests that, assuming a wind fleet of 75 GWc, there would still be an average gas turbine generation requirement of 11.6 GWe to power the grid plus 35 million BEV fleet. The utilisation of the 47.0 GWe gas turbine fleet would therefore be 24.6%.

The National Grid's "System Transformation" scenario for 2050 suggests a residential demand of 39.7 GWe for heating, lighting and cooking, or about 1.5 kW per household (National Grid, 2020). Around 80% of this demand, it suggests, would be provided by steam reforming methane to hydrogen, and using the hydrogen in the homes. Given that wind is to power the BEV fleet, the question arises as to why the residential power needs to be provided by hydrogen. The suggested hydrogen route would require the reforming of methane and, it is argued by the National Grid, that the resulting carbon dioxide could be captured.

However, the proposal to employ carbon capture is a highly controversial one, with many scientists and engineers sceptical about whether the technology is a practical one, particularly at the scale required. Critics of proposals to use carbon capture include the Economist, which describes it as:

> *"a technology of prevarication…. easier, cheaper and more definitive just to generate electricity by other means such as renewables and nuclear".* (The Economist, 2020b)

Taking the power for residences from the grid, courtesy of an increased wind fleet, would reduce the overall need for methane and remove the significant technical risk associated with carbon capture. Adding the residential demand to the grid would also increase the efficiency of the wind fleet and increase the size of the wind fleet which might be economically deployed.

The European Union envisages an investment in 50 to 75 GW of wind and solar capacity dedicated to hydrogen production by 2030 (Sullivan & Cromwell, 2020). With the enormous stored energy of a BEV fleet at its disposal, perhaps the UK would be better advised to put the output of the generating capacity needed to power residences into its grid, obviating the need hydrolyse water, transport, store and eventually burn hydrogen. The BEV owners involved in the OVO Energy trial are already buying the electricity for both their residences and BEVs when the price is low, and selling any surplus back to the grid when the price is high (Lewington, 2020). A household with a single BEV would provide a demand on the grid of around 2 kW (about 1.5 kW for the residence and 0.5 kW for the BEV), and it should be possible for a 30 kWh BEV to manage this time-varying demand to the satisfaction of both BEV owner and the grid.

Although Table 2 reveals that the cost of the battery storage for a 35 million BEV fleet would be about Euro 268 Bn, or nearly Euro 8,000 per BEV (including V2G charger), the cost is borne on purchasing the BEV. Its additional use of smoothing Grid Demand is free of charge, and the variation of price during the day ensures that the benefit of the stored energy is shared by the BEV owner and the grid. It is noted, however, that the 1,050 GWh stored energy of the 35 million BEV fleet, although large, is still quite small compared with the deficits of over 10,000 GWh which are a consequence of lengthy winter wind lulls; a shadow fleet of gas turbines will therefore always be required to mitigate wind lulls.

**Conclusion**

The United Kingdom is considering two major socio-technical transitions, one in the energy sector and the other in the mobility sector. A successful energy transition, including the realisation of a zero-carbon status by 2050, will depend on an increased use of solar and wind energy, despite its intermittent nature. Wind energy, in particular, can be highly unreliable, with the output of wind fleets being a fraction of the total installed capacity for extended periods. Although an expensive option, this problem could be mitigated through the use of a shadow fleet of CCGTs running on natural gas, with a capacity almost equivalent to the total Grid Demand.

However, it may also be possible to use the batteries of a fleet of BEVs, configured as V2G, to offset at least 40% of the required CCGT capacity. Using a mathematical model based on real-time data, our studies have shown that with careful planning, the transitions in the energy and mobility sectors can be aligned so as to reduce the cost of implementation in both systems. By adopting V2G, vehicle owners will be able to offset their operating costs through selling energy to the grid during times of low wind, and the National Grid will reduce the overall capital cost by using the storage capacity of the BEV fleet.

Since it is anticipated that 80% of BEV owners will be able to park the vehicles at their residences, widespread V2G will enable both the powering of residences when supply from the grid is constrained and the charging of BEVs when supply is in excess.  The model shows that this configuration will maintain a constant load on the grid and avoid the use of either expensive alternative storage or hydrogen obtained by reforming methane.